\documentclass[12pt,a4paper]{article}

\usepackage{amsmath}
\usepackage{graphicx}
\usepackage{amsfonts}
\usepackage{xcolor}
\usepackage{cancel}

\begin{document}

\title{Electric charge in hyperbolic motion: the special conformal transformation solution}
\author{C\u alin Galeriu}

\maketitle

\section*{Abstract}

{\it As a simple application of special conformal transformations, 
we derive the electromagnetic field produced by an electric charge in hyperbolic motion.
Unlike other purely algebraic derivations, here we develop a more intuitive
geometrical description, based on Minkowski diagrams.}

\section{Introduction}

Just a few years after the birth of the special relativity theory, Cunningham \cite{cunningham} and Bateman \cite{bateman1, bateman2} discovered that 
Maxwell's equations are also invariant under conformal transformations. 
The conformal group is the largest 
group that conserves null line elements, and consists of space-time translations, 
proper and improper homogeneous Lorentz transformations, 
dilatation (or scale) transformations, 
and special conformal (or acceleration) transformations.
Special conformal transformations are described by the formula:
\begin{equation}
x'^\mu = \frac{x^\mu + b^\mu x^2}{1 + 2 b \cdot x + b^2 x^2},
\label{eq:SCT}
\end{equation}
where $b \cdot x = b^\alpha x_\alpha$, $b^2 = b^\alpha b_\alpha$, and $x^2 = x^\alpha x_\alpha$. 
A special conformal transformation can be decomposed into an inversion in the origin, followed by a translation by a constant 
vector $b^\mu$, and then a second inversion in the new origin.
The inverse transformation of (\ref{eq:SCT}) has the translation in the opposite direction:
\begin{equation}
x^\mu = \frac{x'^\mu - b^\mu x'^2}{1 - 2 b \cdot x' + b^2 x'^2}.
\label{eq:SCTinv}
\end{equation}
The denominator in (\ref{eq:SCT}) or (\ref{eq:SCTinv}) may vanish.
Transformation (\ref{eq:SCT}) has singularity points on
a lightcone with vertex at $v^\mu = - \frac{b^\mu}{b^2}$. 
Transformation (\ref{eq:SCTinv}) 
has singularity points on a lightcone with vertex at $v'^\mu = \frac{b^\mu}{b^2}$.

The original work of Cunningham and Bateman dealt with inversion operations.
After it was discovered that special conformal transformations can transform a straight line (a particle at rest) 
into the two branches of a hyperbola (two particles in hyperbolic motion), this result was used to derive the total electromagnetic field 
produced by two such particles (the so called Born solution) \cite{CodirlaOsborn}. 
It is quite surprising how one particle can transform into two particles, 
in the presence of singularity points on its worldline.
The Born solution itself is strange, because it adds together a retarded field (produced by the particle on one branch of the hyperbola) with an
advanced field (produced by the particle on the other branch of the hyperbola). 
These conceptual difficulties have, most likely, prevented the
effective use of special conformal transformations in undergraduate and graduate classical electrodynamics textbooks.  
Due to the fundamental role of symmetry in theoretical physics, 
and the importance of hyperbolic motion in special relativity,
we believe that a more accessible treatment of 
special conformal transformations is very much needed.
 
The conceptual difficulties mentioned above can be circumvented if, 
instead of transforming one straight line into the two branches of a hyperbola, we transform
just one branch of the hyperbola into one straight line segment. 
In other words, we use the inverse of the special conformal transformation 
that was previously used by other authors. 
This method allows us to derive the electromagnetic field produced by just one electric charge in hyperbolic motion, not the Born solution,
as a simple application of special conformal transformations.

\section{Electric charge in hyperbolic motion}

The electromagnetic field produced by an electric charge in hyperbolic motion was calculated a long time ago by Sommerfeld \cite{galeriuHYP}, based on the retarded electromagnetic four-potential of Li\'{e}nard and Wiechert.
His geometrical derivation takes place in Minkowski space, where a point has coordinates $(x, y, z, i c t)$.
Here we obtain the same electromagnetic field, this time using special conformal transformations.
To facilitate the comparison of results, and to give our intuition the benefit of a geometrical description, 
we also use Minkowski diagrams in our derivation. The same Minkowski metric 
applies to both Minkowski space and flat conformal space.

\begin{figure}[h!]
\begin{center}
\includegraphics[height=10.5cm]{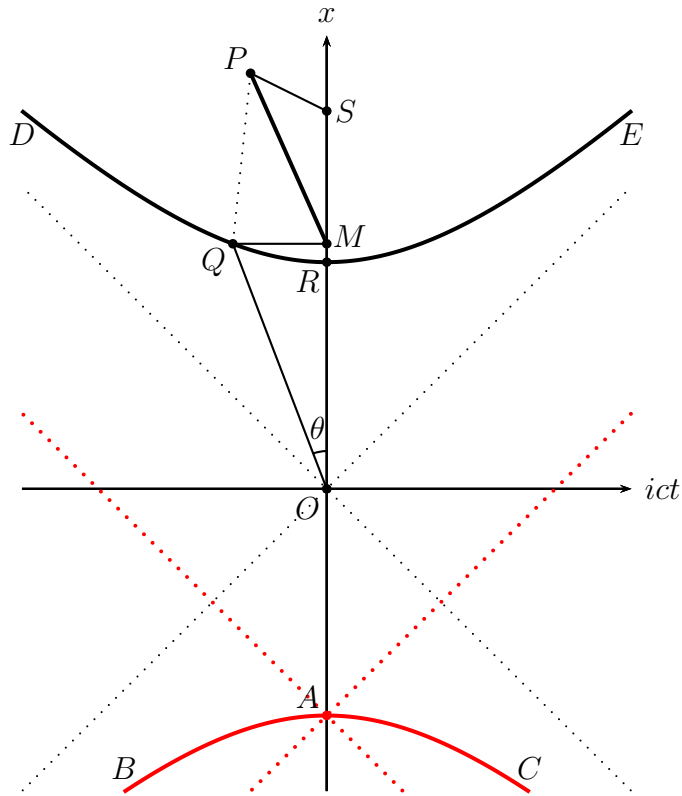}
\caption{The electric field at point $P$, produced by the source charge at $Q$, has the direction of ray $MP$.
This Minkowski diagram assumes that $OR < OS$, where $OR = a$ and $OS = \rho$.
Point $A$ is the vertex of the lightcone of singularity points, and $OA = a$.}
\label{fig:1}
\end{center}
\end{figure}

Consider a source particle with electric charge $e$ in hyperbolic motion, with a trajectory in the $(x, i c t)$ plane.
The center $O$ of the hyperbola coincides with the origin of the reference frame.
The source particle is on the hyperbola branch with positive $x$ coordinates, as shown in Figure \ref{fig:1}.
A point $Q$ on the worldline of the particle has coordinates $X_Q = (a \cos(\psi), 0, 0, a \sin(\psi))$,
where $\psi$ is an imaginary angle measured clockwise from the $Ox$ axis to the $OQ$ ray.
To the left of the $x$ axis angle $\psi$ is negative, and to the right of the $x$ axis angle $\psi$ is positive.
We also define $\theta = - \psi$, 
where $\theta$ is an imaginary angle measured counterclockwise from the $Ox$ axis to the $OQ$ ray.

Consider a test particle at point $P$, with coordinates $X_P = (\rho, y, z, 0)$.
We work in a reference frame in which the test particle at $P$ and the center $O$ of the hyperbola are simultaneous.
If necessary, a Lorentz boost can always bring us to such a reference frame.
The projection of point $P$ onto the $(x, i c t)$ plane is point S, with coordinates $X_S = (\rho, 0, 0, 0)$.

The position four-vector from $Q$ to $P$ is $X = X_P - X_Q$.
We want point $P$ to be on the future light-cone of point $Q$.
Due to this retardation condition, angle $\psi$ is negative and angle $\theta$ is positive.
Because points $Q$ and $P$ are connected by a light signal, we know that $X \cdot X = 0$.
As a result:
\begin{equation}
\cos(\theta) = \frac{\rho^2 + a^2 + y^2 + z^2}{2 a \rho}.
\label{eq:1}
\end{equation}

The electric field produced by the source charge from $Q$ at point $P$ is \cite{galeriuHYP}:
\begin{equation}
E_x = \frac{e \rho ( \rho - a \cos(\theta) )}{a r^3},
\label{eq:2x}
\end{equation}
\begin{equation}
E_y = \frac{e \rho y}{a r^3},
\label{eq:2y}
\end{equation}
\begin{equation}
E_z = \frac{e \rho z}{a r^3},
\label{eq:2z}
\end{equation}
where 
\begin{equation}
r = - i \rho \sin(\theta),
\label{eq:r}
\end{equation}
is the distance to point $P$, measured in a reference frame co-moving
with the source particle at $Q$. The magnitude of the electric field is:
\begin{equation}
|\vec{E}| = \sqrt{E_x^2 + E_y^2 + E_z^2} = \frac{e}{r^2}.
\label{eq:3}
\end{equation}
The magnetic field, in the reference frame considered, is zero.

In order to establish the connection between the electromagnetic field of a charge in hyperbolic motion
and the electromagnetic field of a charge at rest, we start by eliminating any reference to $\theta$ in the above formulas.
Substitution of (\ref{eq:1}) into (\ref{eq:2x}) gives:
\begin{equation}
E_x = \frac{e (\rho^2 - a^2 - y^2 - z^2)}{2 a r^3},
\label{eq:2x_new}
\end{equation}
and the electric field takes the form:
\begin{equation}
\vec{E} = \frac{e}{r^3} \left( \frac{\rho^2 - a^2 - y^2 - z^2}{2 a}, \frac{\rho y}{a}, \frac{\rho z}{a} \right).
\label{eq:E_field_hyp}
\end{equation}
The square of the radial distance $r$ takes the form:
\begin{multline}
r^2 = - \rho^2 \sin^2(\theta) = \rho^2 ( \cos^2(\theta) - 1 ) 
= \rho^2 \left( \frac{\rho^2 + a^2 + y^2 + z^2}{2 a \rho} \right)^2 - \rho^2 \\
= \frac{\rho^4 + a^4 + y^4 + z^4 - 2 \rho^2 a^2 +2 \rho^2 y^2 + 2 \rho^2 z^2 + 2 a^2 y^2 + 2 a^2 z^2 + 2 y^2 z^2}{4 a^2}.
\label{eq:r2}
\end{multline}

\section{The transformation of the coordinates}

We perform the special conformal transformation that has a four-vector $b = (\frac{1}{a}, 0, 0, 0)$. 
As a result we have $b^2 = \frac{1}{a^2}$.

A point $Q$ on the worldline of the source particle has coordinates:
\begin{equation}
X_Q = (a \cos(\psi), 0, 0, a \sin(\psi)),
\label{eq:Q}
\end{equation}
where $\psi$ is an imaginary angle that goes from $-i\infty$ to $i\infty$.
In equation (\ref{eq:SCT}) we substitute $x = X_Q$, and we calculate $x^2 = a^2$,  
$b \cdot x = \cos(\psi)$, and $1 + 2 b \cdot x + b^2 x^2 = 2(1 + \cos(\psi))$. 
After the special conformal transformation, point $Q'$ has coordinates:
\begin{equation}
X_{Q'} = \left( \frac{a}{2}, 0, 0, \frac{a \sin(\psi)}{2(1 + \cos(\psi))} \right).
\label{eq:Qprime}
\end{equation}
We notice that the spatial coordinate is constant, and that the temporal coordinate $ict'$ goes from 
an initial value given by:
\begin{equation}
\lim_{\psi \to -i\infty} \frac{a \sin(\psi)}{2(1 + \cos(\psi))}
= \frac{i a}{2} \lim_{\omega \to -\infty} \frac{\sinh(\omega)}{1 + \cosh(\omega)}
= -\frac{i a}{2},
\label{ictmin}
\end{equation}
to a final value given by:
\begin{equation}
\lim_{\psi \to i\infty} \frac{a \sin(\psi)}{2(1 + \cos(\psi))}
= \frac{i a}{2} \lim_{\omega \to \infty} \frac{\sinh(\omega)}{1 + \cosh(\omega)}
= \frac{i a}{2},
\label{ictmax}
\end{equation}
where $\psi = i \omega$ and $\omega$ is a real number.

In particular, when $\psi = 0$, point $R$ has coordinates:
\begin{equation}
X_R = (a , 0, 0, 0),
\label{eq:R}
\end{equation}
and, after the special conformal transformation, point $R'$ has coordinates:
\begin{equation}
X_{R'} = \left( \frac{a}{2}, 0, 0, 0 \right).
\label{eq:Rprime}
\end{equation}

Point $P$ has coordinates:
\begin{equation}
X_P = (\rho, y, z, 0).
\label{eq:P}
\end{equation}
In equation (\ref{eq:SCT}) we substitute $x = X_P$, and we calculate $x^2 = \rho^2 + y^2 + z^2$,  
$b \cdot x = \frac{\rho}{a}$, and 
$1 + 2 b \cdot x + b^2 x^2 = (1 + \frac{\rho}{a})^2 + \frac{y^2 + z^2}{a^2} \equiv \sigma$. 
We notice that $\sigma$ is a positive real number.
After the special conformal transformation, point $P'$ has coordinates:
\begin{equation}
X_{P'} = \left( \frac{a \rho (a + \rho) + a (y^2 + z^2) }{(a + \rho)^2 + y^2 + z^2}, 
\frac{a^2 y}{(a + \rho)^2 + y^2 + z^2}, \frac{a^2 z}{(a + \rho)^2 + y^2 + z^2}, 0 \right).
\label{eq:Pprime}
\end{equation}

In particular, when $y = 0$ and $x = 0$, point $S$ has coordinates:
\begin{equation}
X_S = (\rho, 0, 0, 0),
\label{eq:S}
\end{equation}
and, after the special conformal transformation, point $S'$ has coordinates:
\begin{equation}
X_{S'} = \left( \frac{a \rho}{a + \rho}, 0, 0, 0 \right),
\label{eq:Sprime}
\end{equation}
which also means that point $S'$ is no longer the projection of point $P'$.

The special conformal transformation preserves the ordering of points $S'$ and $R'$ on the $O x'$ axis. 
If $\rho > a$ then $\frac{a \rho}{a + \rho} > \frac{a}{2}$, and if $\rho < a$ then $\frac{a \rho}{a + \rho} < \frac{a}{2}$.

\section{Electric charge at rest}

\begin{figure}[h!]
\begin{center}
\includegraphics[height=8.5cm]{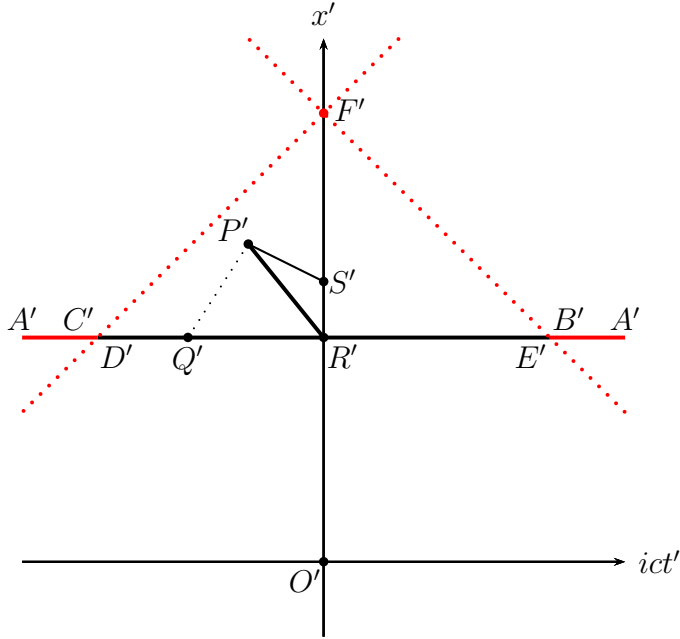}
\caption{The electric field at point $P'$, produced by the source charge at $Q'$, has the direction of ray $R'P'$.
This Minkowski diagram assumes that $O'R' < O'S'$, where $O'R' = \frac{a}{2}$ and $O'S' = \frac{a \rho}{a + \rho}$.
Point $F'$ is the vertex of the lightcone of singularity points, and $O'F' = a$.}
\label{fig:2}
\end{center}
\end{figure}

Consider a source particle with electric charge $e$ at rest, with a worldline given by 
$X_{Q'} = (\frac{a}{2}, 0, 0, i c t')$, as shown in Figure \ref{fig:2}.
The calculation of the electric field at $P'$ is not affected by our assumption that $- i a / 2 < i c t' < i a / 2$.
When the time $t'$ is zero the source particle has coordinates $X_{R'}$ given by (\ref{eq:Rprime}),
and the test particle has coordinates $X_{P'}$ given by (\ref{eq:Pprime}) or by (\ref{eq:Sprime}).

Since:
\begin{equation}
\frac{a \rho (a + \rho) + a (y^2 + z^2) }{(a + \rho)^2 + y^2 + z^2} - \frac{a}{2} = 
\frac{a}{2} \frac{\rho^2 - a^2 + y^2 + z^2}{(a + \rho)^2 + y^2 + z^2},
\label{eq:delta_x}
\end{equation}
the position four-vector $X_{R'P'} = X_{P'} - X_{R'}$, from point $R'$ to point $P'$, is:
\begin{equation}
X_{R'P'} = \left( \frac{a}{2} \frac{\rho^2 - a^2 + y^2 + z^2}{(a + \rho)^2 + y^2 + z^2}, 
\frac{a^2 y}{(a + \rho)^2 + y^2 + z^2}, \frac{a^2 z}{(a + \rho)^2 + y^2 + z^2}, 0 \right).
\label{eq:PprimeRprime}
\end{equation}

The square of the distance from point $R'$ to point $P'$ is:
\begin{multline}
r'^2 = (R'P')^2 = \left( \frac{a}{2} \right)^2 \left( \frac{\rho^2 - a^2 + y^2 + z^2}{(a + \rho)^2 + y^2 + z^2} \right)^2 
+ \frac{a^4 (y^2 + z^2)}{[(a + \rho)^2 + y^2 + z^2]^2} \\
= \frac{a^2}{4} \frac{\rho^4 + a^4 + y^4 + z^4 - 2 \rho^2 a^2 +2 \rho^2 y^2 + 2 \rho^2 z^2 + 2 a^2 y^2 + 2 a^2 z^2 + 2 y^2 z^2}
{[(a + \rho)^2 + y^2 + z^2]^2} \\
= \frac{a^4 r^2}{[(a + \rho)^2 + y^2 + z^2]^2} = \frac{r^2}{\sigma^2}.
\label{eq:RPprime}
\end{multline}
The Coulomb electric field produced at point $P'$ by the source charge at rest at point $R'$ is:
\begin{equation}
E'_{x'} = \frac{e}{r'^3} \frac{a}{2} \frac{\rho^2 - a^2 + y^2 + z^2}{(a + \rho)^2 + y^2 + z^2}
= \frac{e}{r'^3} \frac{\rho^2 - a^2 + y^2 + z^2}{2 a \sigma},
\label{eq:6x}
\end{equation}
\begin{equation}
E'_{y'} = \frac{e}{r'^3} \frac{a^2 y}{(a + \rho)^2 + y^2 + z^2} = \frac{e}{r'^3} \frac{y}{\sigma},
\label{eq:6y}
\end{equation}
\begin{equation}
E'_{z'} = \frac{e}{r'^3} \frac{a^2 z}{(a + \rho)^2 + y^2 + z^2} = \frac{e}{r'^3} \frac{z}{\sigma},
\label{eq:6z}
\end{equation}
which we can also write in the equivalent form:
\begin{equation}
\vec{E'} = \frac{e}{r'^3} \left( \frac{\rho^2 - a^2 + y^2 + z^2}{2 a \sigma}, \frac{y}{\sigma}, \frac{z}{\sigma} \right).
\label{eq:E_field_rest}
\end{equation}
The magnitude of the electric field is:
\begin{equation}
|\vec{E'}| = \sqrt{{E'_{x'}}^2 + {E'_{y'}}^2 + {E'_{z'}}^2} = \frac{e}{r'^2}.
\label{eq:mag_E_prime}
\end{equation}

\section{The transformation of the electric field}

In Minkowski space, when the magnetic field is zero, as it is in the two reference frames considered, 
the electromagnetic field strength tensor takes the form:
\begin{equation}
\mathbb{F} = || F_{\alpha \beta} || = || F^{\alpha \beta} || = 
\begin{pmatrix} 
0 & 0 & 0 & - i E_x \\
0 & 0 & 0 & - i E_y \\
0 & 0 & 0 & - i E_z \\
i E_x & i E_y & i E_z & 0 \end{pmatrix}.
\label{eq:Fmatrix}
\end{equation}

We introduce the matrices
\begin{equation}
\mathbb{M} = || M^\alpha_{\ \beta} || = \left| \left| \frac{\partial x'^\alpha}{\partial x^\beta} \right| \right|,
\label{eq:Mmatrix}
\end{equation}
\begin{equation}
\mathbb{W} = || W^\alpha_{\ \beta} || = \left| \left| \frac{\partial x^\alpha}{\partial x'^\beta} \right| \right|,
\label{eq:Wmatrix}
\end{equation}
where $\alpha$ is a row index, and $\beta$ is a column index. 

As discussed by Nicholas Wheeler \cite{NicholasWheeler},
under the special conformal transformation (\ref{eq:SCT}) the covariant $F_{\alpha \beta}$ transforms like a tensor:
\begin{equation}
F'_{\mu \nu}(x') = 
\frac{\partial x^\alpha}{\partial x'^\mu} \frac{\partial x^\beta}{\partial x'^\nu} F_{\alpha \beta}(x),
\label{eq:Fcovariant}
\end{equation}
and the contravariant $F^{\alpha \beta}$ transforms like a tensor density of weight 1:
\begin{equation}
F'^{\mu \nu}(x') = W
\frac{\partial x'^\mu}{\partial x^\alpha} \frac{\partial x'^\nu}{\partial x^\beta} F^{\alpha \beta}(x),
\label{eq:Fcontravariant}
\end{equation}
where $W$ is the Jacobian determinant of the transformation $x \mapsto x'$ :
\begin{equation}
W = | \mathbb{W} |
= \frac{1}{(1 - 2 b \cdot x' + b^2 x'^2)^4} 
= (1 + 2 b \cdot x + b^2 x^2)^4 = \sigma^4.
\label{eq:Jacobian}
\end{equation}
A similar set of equations is obtained for the inverse special conformal transformation (\ref{eq:SCTinv}). 
Equation (\ref{eq:Fcovariant}) becomes:
\begin{equation}
F_{\mu \nu}(x) = 
\frac{\partial x'^\alpha}{\partial x^\mu} \frac{\partial x'^\beta}{\partial x^\nu} F'_{\alpha \beta}(x'),
\label{eq:Fcovariantinv}
\end{equation}
and equation (\ref{eq:Fcontravariant}) becomes:
\begin{equation}
F^{\mu \nu}(x) = M
\frac{\partial x^\mu}{\partial x'^\alpha} \frac{\partial x^\nu}{\partial x'^\beta} F'^{\alpha \beta}(x'),
\label{eq:Fcontravariantinv}
\end{equation}
where $M$ is the Jacobian determinant of the transformation $x' \mapsto x$ :
\begin{equation}
M = | \mathbb{M} |
= \frac{1}{(1 + 2 b \cdot x + b^2 x^2)^4} = \frac{1}{\sigma^4}.
\label{eq:Jacobianinv}
\end{equation}

In flat conformal space the Lorentz invariant $F_{\alpha \beta} F^{\alpha \beta}$ transforms like a scalar density of weight 1:
\begin{equation}
F'_{\mu \nu}(x') F'^{\mu \nu}(x') = W F_{\alpha \beta}(x) F^{\alpha \beta}(x).
\label{eq:FF}
\end{equation}
Since in the two reference frames considered the magnetic field is zero, 
$F_{\alpha \beta} F^{\alpha \beta} = - 2 \vec{E}^2 = - 2 e^2 / r^4$,
$F'_{\mu \nu} F'^{\mu \nu} = - 2 \vec{E'}^2 = - 2 e^2 / r'^4$,
and equation (\ref{eq:FF}) provides an alternative derivation of the equation (\ref{eq:RPprime}), $r' = r / \sigma$.

From equation (\ref{eq:SCT}) we calculate \cite{Ryder1974}:
\begin{equation}
M^\mu_{\ \alpha} = \frac{\partial x'^\mu}{\partial x^\alpha} 
= \frac{\delta^\mu_{\ \alpha} + 2 b^\mu x_\alpha}{1 + 2 b \cdot x + b^2 x^2} 
- \frac{(x^\mu + b^\mu x^2)(2 b_\alpha + 2 x_\alpha b^2)}{(1 + 2 b \cdot x + b^2 x^2)^2},
\label{eq:dxpmdxa}
\end{equation}
and from equation (\ref{eq:SCTinv}) we calculate:
\begin{equation}
W^\alpha_{\ \mu} = \frac{\partial x^\alpha}{\partial x'^\mu} 
= \frac{\delta^\alpha_{\ \mu} - 2 b^\alpha x'_\mu}{1 - 2 b \cdot x' + b^2 x'^2} 
- \frac{(x'^\alpha - b^\alpha x'^2)(- 2 b_\mu + 2 x'_\mu b^2)}{(1 - 2 b \cdot x' + b^2 x'^2)^2}.
\label{eq:dxadxpm}
\end{equation}

The equations (\ref{eq:Fcovariant}), (\ref{eq:Fcontravariant}), (\ref{eq:Fcovariantinv}), (\ref{eq:Fcontravariantinv}) 
can be used to calculate one electric field from the other one,
and this is most easily done by writing these equations in matrix form:
\begin{equation}
\mathbb{F}' = \mathbb{W}^T \ \mathbb{F} \ \mathbb{W},
\label{eq:WTFW}
\end{equation}
\begin{equation}
\mathbb{F}' = W \ \mathbb{M} \ \mathbb{F} \ \mathbb{M}^T,
\label{eq:MFMT}
\end{equation}
\begin{equation}
\mathbb{F} = \mathbb{M}^T \ \mathbb{F}' \ \mathbb{M},
\label{eq:MTFM}
\end{equation}
\begin{equation}
\mathbb{F} = M \ \mathbb{W} \ \mathbb{F}' \ \mathbb{W}^T.
\label{eq:WFWT}
\end{equation}
Because the matrices $\mathbb{M}$ and $\mathbb{W}$ are inverses of each other,
it is obvious that equations (\ref{eq:WTFW}) and (\ref{eq:MTFM}) are equivalent, 
and also equations (\ref{eq:MFMT}) and (\ref{eq:WFWT}).

Direct substitution of $b = (\frac{1}{a}, 0, 0, 0)$ and of $x = X_P = (\rho, y, z, 0)$ into equation (\ref{eq:dxpmdxa}) gives:
\begin{equation}
\mathbb{M} = \begin{pmatrix} 
A & B & C & 0 \\
D & E & F & 0 \\
G & H & I & 0 \\
0 & 0 & 0 & J \end{pmatrix},
\label{eq:Mmatrix2}
\end{equation}
where
\begin{multline}
A \equiv \frac{\partial x'^1}{\partial x^1} 
= \left( 1 + 2 \frac{\rho}{a} \right) \frac{1}{\sigma}
- \left( \rho + \frac{\rho^2 + y^2 + z^2}{a} \right) \left( \frac{2}{a} + \frac{2 \rho}{a^2} \right) \frac{1}{\sigma^2} \\
= \frac{1}{\sigma} - \frac{2 ( y^2 + z^2 )}{a^2 \sigma^2}
= \frac{a^2 + 2 a \rho + \rho^2 - y^2 - z^2}{a^2 \sigma^2},
\label{eq:A}
\end{multline}
\begin{equation}
B \equiv \frac{\partial x'^1}{\partial x^2} 
= \frac{2 y}{a} \frac{1}{\sigma} 
- \left( \rho + \frac{\rho^2 + y^2 + z^2}{a} \right) \frac{2 y}{a^2} \frac{1}{\sigma^2}
= \frac{2 y (a + \rho)}{a^2 \sigma^2},
\label{eq:B}
\end{equation}
\begin{equation}
C \equiv \frac{\partial x'^1}{\partial x^3} 
= \frac{2 z}{a} \frac{1}{\sigma} 
- \left( \rho + \frac{\rho^2 + y^2 + z^2}{a} \right) \frac{2 z}{a^2} \frac{1}{\sigma^2}
= \frac{2 z (a + \rho)}{a^2 \sigma^2},
\label{eq:C}
\end{equation}
\begin{equation}
D \equiv \frac{\partial x'^2}{\partial x^1} 
= - y \left( \frac{2}{a} + \frac{2 \rho}{a^2} \right) \frac{1}{\sigma^2}
= - \frac{2 y (a + \rho)}{a^2 \sigma^2},
\label{eq:D}
\end{equation}
\begin{equation}
E \equiv \frac{\partial x'^2}{\partial x^2} 
= \frac{1}{\sigma} - \frac{2 y^2}{a^2 \sigma^2}
= \frac{a^2 + 2 a \rho + \rho^2 - y^2 + z^2}{a^2 \sigma^2},
\label{eq:E}
\end{equation}
\begin{equation}
F \equiv \frac{\partial x'^2}{\partial x^3} 
= - \frac{2 y z}{a^2 \sigma^2},
\label{eq:F}
\end{equation}
\begin{equation}
G \equiv \frac{\partial x'^3}{\partial x^1} 
= - z \left( \frac{2}{a} + \frac{2 \rho}{a^2} \right) \frac{1}{\sigma^2}
= - \frac{2 z (a + \rho)}{a^2 \sigma^2},
\label{eq:G}
\end{equation}
\begin{equation}
H \equiv \frac{\partial x'^3}{\partial x^2} 
= - \frac{2 y z}{a^2 \sigma^2},
\label{eq:H}
\end{equation}
\begin{equation}
I \equiv \frac{\partial x'^3}{\partial x^3} 
= \frac{1}{\sigma} - \frac{2 z^2}{a^2 \sigma^2}
= \frac{a^2 + 2 a \rho + \rho^2 + y^2 - z^2}{a^2 \sigma^2},
\label{eq:I}
\end{equation}
\begin{equation}
J \equiv \frac{\partial x'^4}{\partial x^4}
= \frac{1}{\sigma}. 
\label{eq:J}
\end{equation}

Direct substitution of matrix (\ref{eq:Mmatrix2}) into equation (\ref{eq:MFMT}) gives the electric field $\vec{E'}$ as 
a function of the electric field $\vec{E}$:
\begin{equation}
E'_{x'} = W ( A E_x + B E_y + C E_z ) J,
\label{eq:ExpABC}
\end{equation}
\begin{equation}
E'_{y'} = W ( D E_x + E E_y + F E_z ) J,
\label{eq:EypDEF}
\end{equation}
\begin{equation}
E'_{z'} = W ( G E_x + H E_y + I E_z ) J.
\label{eq:EzpGHI}
\end{equation}
The substitution of the coefficients (\ref{eq:A})-(\ref{eq:J}),
of the Jacobian determinant (\ref{eq:Jacobian}),
and of the electric field (\ref{eq:E_field_hyp}) into equations (\ref{eq:ExpABC})-(\ref{eq:EzpGHI})
produces, after a simple but somewhat lengthy algebraic calculation, the electric field (\ref{eq:E_field_rest}). 
During this calculation we use the relations $r = \sigma r'$ and $a^2 \sigma = a^2 + 2 a \rho + \rho^2 + y^2 + z^2$.
For the calculation of $E'_{x'}$ we also need the factorization:
\begin{multline}
\rho^4 - a^4 + y^4 + z^4 + 2 a \rho^3 - 2 a^3 \rho + 2 \rho^2 y^2 + 2 \rho^2 z^2 + 2 y^2 z^2 + 2 a \rho y^2 + 2 a \rho z^2 \\
= ( \rho^2 - a^2 + y^2 + z^2 ) ( a^2 + 2 a \rho + \rho^2 + y^2 + z^2 ).
\label{eq:factorization}
\end{multline}
For example, the calculation of $E'_{z'}$ proceeds as follows:
\begin{multline}
E'_{z'} = W ( G E_x + H E_y + I E_z ) J
= \sigma^4 \bigg(
\frac{- \bcancel{2} z (a + \rho)}{a^2 \sigma^2} \frac{e}{r^3} \frac{(\rho^2 - a^2 - y^2 - z^2)}{\bcancel{2} a} \\
+ \frac{(- 2 y z)}{a^2 \sigma^2} \frac{e}{r^3} \frac{\rho y}{a}
+ \frac{(a^2 + 2 a \rho + \rho^2 + y^2 - z^2)}{a^2 \sigma^2} \frac{e}{r^3} \frac{\rho z}{a} 
\bigg) \frac{1}{\sigma} \\
= \frac{e}{r'^3} \bigg(
\frac{- a \rho^2 z + a^3 z + a y^2 z + a z^3 - \cancel{\rho^3 z} + a^2 \rho z + \xcancel{\rho y^2 z} + \bcancel{\rho z^3}}{a^3 \sigma^2} \\
- \frac{\xcancel{2 \rho y^2 z}}{a^3 \sigma^2} 
+ \frac{a^2 \rho z + 2 a \rho^2 z + \cancel{\rho^3 z} + \xcancel{\rho y^2 z} - \bcancel{\rho z^3}}{a^3 \sigma^2} \bigg) \\
= \frac{e}{r'^3} \frac{a^3 z + 2 a^2 \rho z + a \rho^2 z + a y^2 z + a z^3}{a \sigma (a^2 + 2 a \rho + \rho^2 + y^2 + z^2)}
= \frac{e}{r'^3} \frac{z}{\sigma}.
\label{eq:EzpGHIcalc}
\end{multline}

Direct substitution of matrix (\ref{eq:Mmatrix2}) into equation (\ref{eq:MTFM}) gives the electric field $\vec{E}$ as 
a function of the electric field $\vec{E'}$:
\begin{equation}
E_x = ( A E'_{x'} + D E'_{y'} + G E'_{z'} ) J,
\label{eq:ExADG}
\end{equation}
\begin{equation}
E_y = ( B E'_{x'} + E E'_{y'} + H E'_{z'} ) J,
\label{eq:EyBEH}
\end{equation}
\begin{equation}
E_z = ( C E'_{x'} + F E'_{y'} + I E'_{z'} ) J.
\label{eq:EzCFI}
\end{equation}
The substitution of the coefficients (\ref{eq:A})-(\ref{eq:J})
and of the electric field (\ref{eq:E_field_rest}) into equations (\ref{eq:ExADG})-(\ref{eq:EzCFI})
produces, after a simple but somewhat lengthy algebraic calculation, the electric field (\ref{eq:E_field_hyp}). 
For the calculation of $E_x$ we also need the factorization:
\begin{multline}
\rho^4 - a^4 - y^4 - z^4 + 2 a \rho^3 - 2 a^3 \rho - 2 a^2 y^2 - 2 a^2 z^2 - 2 y^2 z^2 - 2 a \rho y^2 - 2 a \rho z^2 \\
= ( \rho^2 - a^2 - y^2 - z^2 ) ( a^2 + 2 a \rho + \rho^2 + y^2 + z^2 ).
\label{eq:factorization2}
\end{multline}

For example, the calculation of $E_z$ proceeds as follows:
\begin{multline}
E_z = ( C E'_{x'} + F E'_{y'} + I E'_{z'} ) J 
= \bigg( 
\frac{\bcancel{2} z (a + \rho)}{a^2 \sigma^2} \frac{e}{r'^3} \frac{(\rho^2 - a^2 + y^2 + z^2)}{\bcancel{2} a \sigma} \\
+ \frac{\textcolor{red}{a \cdot} (- 2 y z)}{\textcolor{red}{a \cdot} a^2 \sigma^2} \frac{e}{r'^3} \frac{y}{\sigma}
+ \frac{\textcolor{red}{a \cdot} (a^2 + 2 a \rho + \rho^2 + y^2 - z^2)}
{\textcolor{red}{a \cdot} a^2 \sigma^2} \frac{e}{r'^3} \frac{z}{\sigma}
\bigg) \frac{1}{\sigma} \\
= \frac{e}{\sigma^3 r'^3} \bigg(
\frac{a \rho^2 z - \cancel{a^3 z} + \xcancel{a y^2 z} + \bcancel{a z^3} + \rho^3 z - a^2 \rho z + \rho y^2 z + \rho z^3}{a^3 \sigma} \\
- \frac{\xcancel{2 a y^2 z}}{a^3 \sigma}
+ \frac{\cancel{a^3 z} + 2 a^2 \rho z + a \rho^2 z + \xcancel{a y^2 z} - \bcancel{a z^3}}{a^3 \sigma} \bigg) \\
= \frac{e}{r^3} \frac{a^2 \rho z + 2 a \rho^2 z + \rho^3 z + \rho y^2 z + \rho z^3}{a (a^2 + 2 a \rho + \rho^2 + y^2 + z^2)}
= \frac{e}{r^3} \frac{\rho z}{a}.
\label{eq:EzCFIcalc}
\end{multline}

As a side note, in the situation when the restrictions
$b = (\frac{1}{a}, 0, 0, 0)$ and of $x = X_P = (\rho, y, z, 0)$
do not apply, substitution of (\ref{eq:dxpmdxa}) into (\ref{eq:Fcontravariant}), or of (\ref{eq:dxadxpm}) into (\ref{eq:Fcovariant}),
produces the most general formula relating the electromagnetic fields
before and after the special conformal transformation.
This formula was also obtained by Barut and Haugen \cite{BarutHaugen1972}
with the help of a six-dimensional linear representation of the conformal group.

\section{Conclusions}

The conformal symmetry of Maxwell's equations remains a topic still avoided by
most electrodynamics textbooks. Some conceptual difficulties seem to bear the blame for this situation.

When we derive the electromagnetic field of an accelerated electric charge,
starting with one electric charge at rest (worldline $A' C' D' Q' R' E' B' A'$ in Figure \ref{fig:2})
and applying the special conformal transformation (\ref{eq:SCTinv}), we notice the 
apparition of two particles (worldlines $D Q R E$ and $B A C$ in Figure \ref{fig:1}).
Fulton and Rohrlich \cite{FultonRohrlich1060}, giving credit to G\"{u}rsey for valuable comments, 
have analyzed this transformation in detail, providing Minkowski diagrams similar to ours.

We also notice the temporal inversion of points B and C, which indicates possible violations of causality.
Rosen \cite{JRosen1960} has analyzed this outcome in detail, showing that the reversal of temporal ordering 
happens only when there is a singularity point on the trajectory between the two spacepoints considered.

The Coulomb electric field of an electric charge at rest fills the whole of spacetime, and it was assumed
that the electric field after the special conformal transformation has the same property. Since the 
electric field produced by just one electric charge in hyperbolic motion does not fill the whole of spacetime
(even if we combine the retarded field with the advanced field), the apparition of the second particle seemed necessary.
Codirla and Osborn \cite{CodirlaOsborn} justify in this way the derivation of the Born solution: 
"The fields obtained by conformal transformation are nonzero
everywhere for all time and are, of course, solutions of Maxwell's equations. They
are related to, but not identical with, the standard retarded, or advanced, solutions,
since these are zero on half of space-time." 

We do not embrace the assumption that the electromagnetic field, after the special conformal transformation,
should fill all of the spacetime, instead we limit our investigation to regions away from the lightcone of singularity points. 
This restriction was made clear by Fulton, Rohrlich, and Witten \cite{FRWcimento}: 
"The fact that the transformation is singular is in no way disturbing. One must merely obey the injunction to stay away from
the singularities in discussing any physical process. Using this transformation, we discuss physical processes only for regions
in space-time for which the transformation is nonsingular." 

In our approach the electric charge at rest is not represented by the infinite worldline $A' C' D' Q' R' E' B' A'$, 
but by the finite worldsegment $D' Q' R' E'$. In this way we avoid the singularity points, with all their related
conceptual difficulties. Although the infinite worldline $A' C' D' Q' R' E' B' A'$ in flat conformal space 
may look like the infinite worldline of a particle at rest in Minkowski space, in truth they are not the same.
This is because the special conformal transformation (\ref{eq:SCTinv}), acting on a particle in hyperbolic motion in Minkowski space,
will change the rest mass of the particle along its worldline. This important fact can be easily overlooked because the
electric charge of a particle, unlike its rest mass, is invariant under conformal transformations.

We also know that the selection of the retarded electromagnetic potential, at the expense of the advanced one,
reduces the symmetry group of classical relativistic physics to the inhomogeneous Lorentz group plus dilatations \cite{Zeeman}.
The conformal symmetry of Maxwell's equations could play an important role in theories of time-symmetric
electrodynamic interaction. In particular, the physical systems described in Figures \ref{fig:1} and \ref{fig:2} exhibit
temporal symmetry. In these two cases the same electric fields are obtained from retarded potentials, from advanced potentials, or from a
linear combination of both.

In conclusion, we hope that our intuitive, geometrical derivation presented here
will become a pedagogical example of how to use the conformal symmetry of Maxwell's equations in order
to derive the electromagnetic field produced by an electric charge in hyperbolic motion.

\end{document}